\documentclass{kluwer}    
\usepackage{graphics,lscape}

\newdisplay{guess}{Conjecture}

\begin{document}                                                                                   
\begin{article}
\begin{opening}         
\title{RX\,J0944.5+0357: A Probable Intermediate Polar\thanks{This paper uses observations
made from the South African Astronomical Observatory (SAAO).}} 
\author{Patrick A. \surname{Woudt} and Brian \surname{Warner}}  
\runningauthor{Patrick A. Woudt and Brian Warner}
\runningtitle{RX\,J0944.5+0357: A Probable Intermediate Polar}
\institute{Dept. of Astronomy, University of Cape Town, Rondebosch 7700, South Africa\\
E-mail: pwoudt@circinus.ast.uct.ac.za, warner@physci.uct.ac.za}
\date{30 October 2002}

\begin{abstract}
From optical photometry the cataclysmic variable RX\,J0944.5+0357 is shown
to have a double-peaked pulse profile with a period $\sim$ 2160 s. The two peaks 
vary rapidly in relative amplitude. Often most of the optical power is concentrated in 
the first harmonic of the 2160 s modulation; RX\,J0944.5+0357 therefore probably belongs 
to the relatively rare class of two-pole accreting intermediate polars exemplified 
by YY Dra and V405 Aur.
\end{abstract}

\end{opening}

\section{Introduction}
Many of the X-Ray sources in the ROSAT All-Sky Survey have been identified 
optically in the Hamburg objective prism survey (Hagen et al. 1995), among which are 
several cataclysmic variables (CVs) (Jiang et al. 2000). The source RX\,J0944.5+0357 
(= 1RXS\,J094432.1+035738; hereafter RXJ0944), in the constellation Sextans, was 
observed spectroscopically by Jiang et al. and found to have H\,I and He\,I emission 
lines typical of a CV. Further spectroscopic study by Mennickent et al. (2002) showed the 
presence of absorption bands in the red, characteristic of a secondary with a spectral 
type near M2.
Observations by the VSNET group have identified two dwarf nova-like 
outbursts, in January and June 2001, during which RXJ0944 rose to V $\sim$ 13 from its 
quiescent magnitude of V $\sim$ 16.2. Mennickent et al. confirmed the spectroscopically 
determined orbital period ($P_{orb}$) of 0.1492 d (3.581 h) reported to them 
by Thorstensen \& Fenton.
Mennickent et al. also provided the first high speed photometry of RXJ0944 in 
which large amplitude variations ($\sim$ 0.5 mag) were found on time scales of 10 min 
to 2 h. They did not report any coherent signals in their photometry.

\section{Photometric Observations}

\begin{figure}
\centering
 \resizebox{\hsize}{!}{\includegraphics{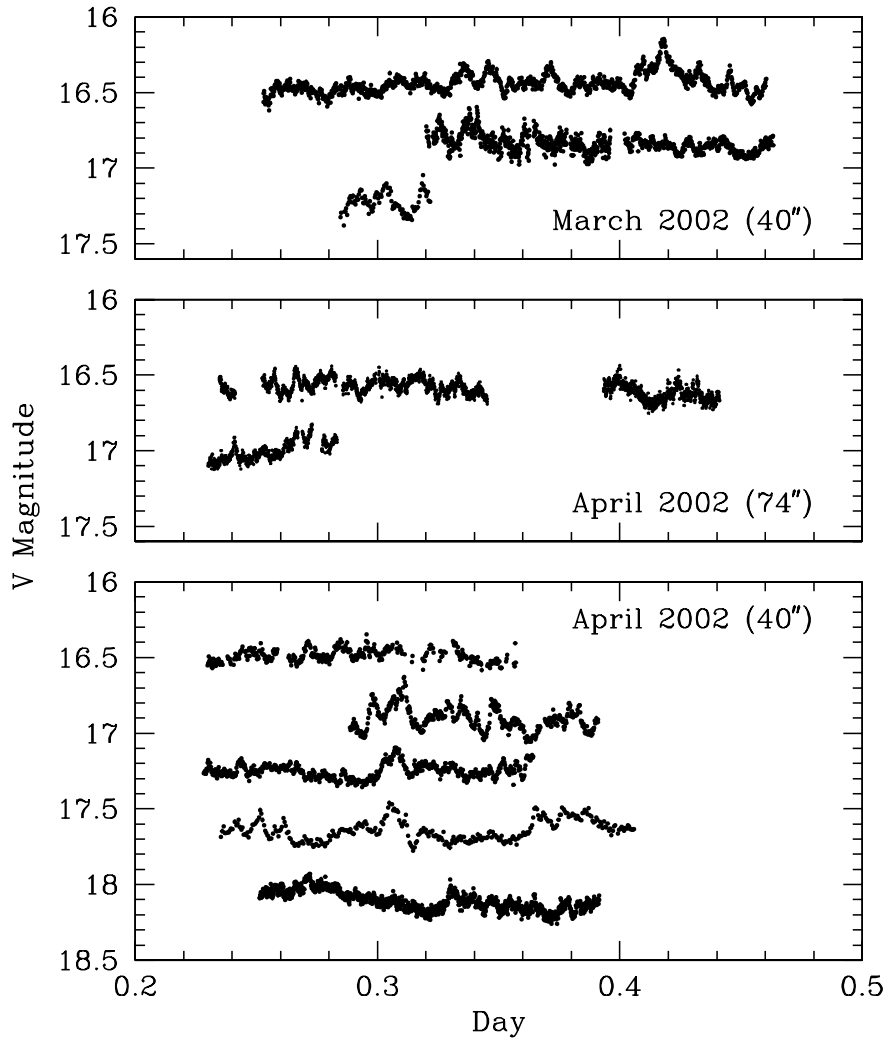}}
 \caption[]{The light curves of RXJ0944, shown in chronological order from top to bottom. In the top panel,
runs S6331 and S6340 are displaced downwards by 0.35 and 0.7 mag, respectively. 
In the middle panel run S6350 is displaced downwards by 0.5 mag. In the lower panel runs S6366, S6370, S6386 and S6391
are displaced vertically by 0.6, 0.75, 1.1 and 1.5 mag, respectively. }
 \label{fig1}
\end{figure}

We have used the University of Cape Town CCD Photometer (O'Dono\-ghue 1995),
attached to the 74-in and 40-in telescopes at the Sutherland site of the 
South African Astronomical Observatory, to observe RXJ0944 at time resolutions 
down to 6 s. Table 1 gives the log of our photometric observations and Figure~\ref{fig1} 
shows the resulting light curves.
    
\begin{table}
 \centering
  \caption{Observing log.}
  \begin{tabular}{@{}ccccccc@{}}
\hline
 Run No.  & Date of obs.          & HJD of first obs. & Length    & $t_{in}$ & Tel.    & $<$V$>$ \\ 
          & (start of night)      &  (+2452000.0)     & (h)       &     (s)   &        & (mag)   \\[10pt]
\hline
 S6324    & 22 March 2002         &    356.25290      &   4.98    &      15   &  40-in &  16.4    \\
 S6331    & 23 March 2002         &    357.40201      &   1.47    &      15   &  40-in &  16.5    \\
 S6340    & 25 March 2002         &    359.28466      &   0.89    &      30   &  40-in &  16.5    \\
 S6341    & 02 April 2002         &    367.23471      &   4.96    &       6   &  74-in &  16.6:   \\
 S6350    & 06 April 2002         &    371.23004      &   1.28    &       8   &  74-in &  16.5:   \\
 S6362    & 09 April 2002         &    374.22997      &   3.05    &      20   &  40-in &  16.5    \\
 S6366    & 10 April 2002         &    375.28856      &   2.46    &      20   &  40-in &  16.3    \\
 S6370    & 11 April 2002         &    376.22836      &   3.26    &      20   &  40-in &  16.5    \\
 S6386    & 14 April 2002         &    379.23542      &   4.09    &      45   &  40-in &  16.5    \\
 S6391    & 15 April 2002         &    380.25135      &   3.36    &      10   &  40-in &  16.5    \\
 \hline
 \end{tabular}
{\footnotesize
\newline
Notes: `:' denotes an uncertain value, $t_{in}$ is the integration time.\hfill}
\label{tab1}
\end{table}

A Fourier Transform (FT) of the entire data set shows no power at the 
spectroscopic period or its first harmonic, so we deduce that RXJ0944 is of quite low 
inclination. From the radial velocity amplitude of 75 km s$^{-1}$ Mennickent et al. 
reasoned that the inclination probably lies in the range $30^{\circ} - 60^{\circ}$; 
our result indicates that it is probably at the lower end
of this range. A low inclination is also compatible with the weakness of the 
emission lines in the spectrum.
     
\begin{figure}
\centering
 \resizebox{\hsize}{!}{\includegraphics{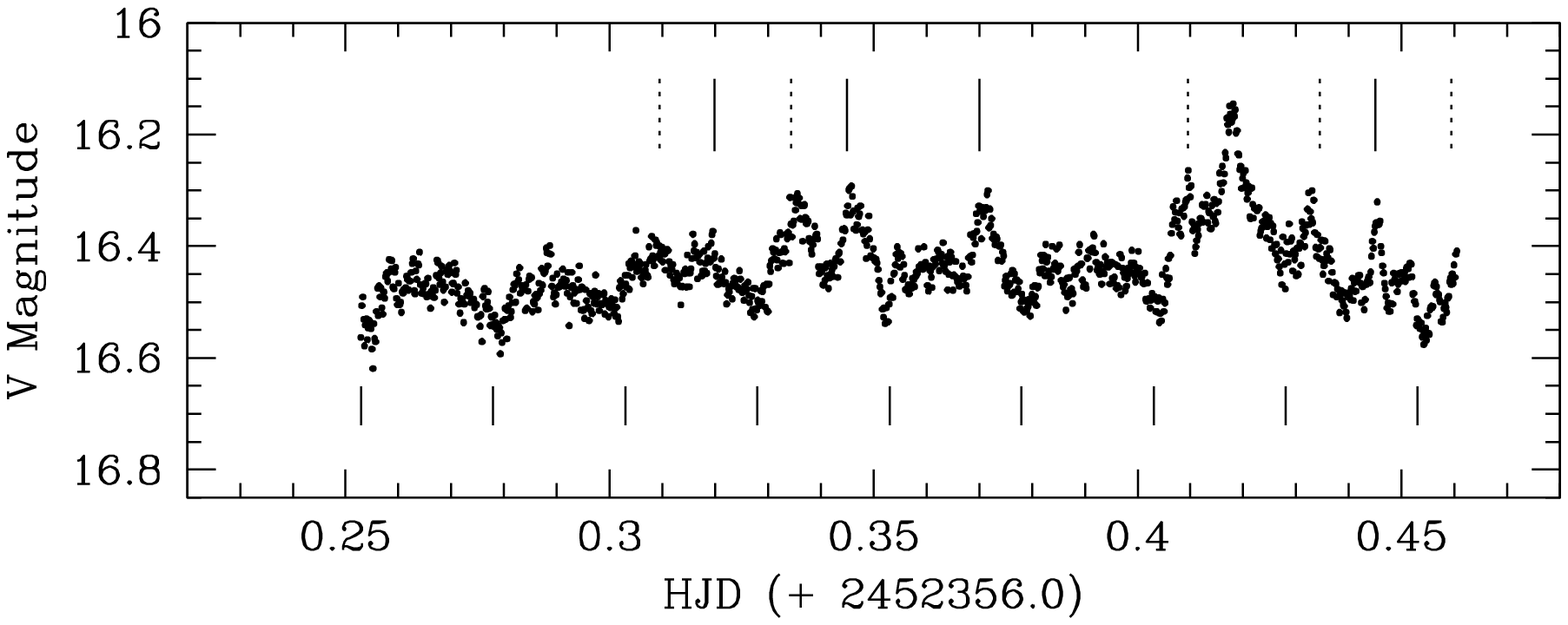}}
 \caption[]{The light curve of RXJ0944 (run S6324). The 2162 s periodicity is marked by the 
lower vertical bars; the pairs of peaks of variable amplitude are marked by the upper bars (dotted bar for
the first peak, solid bar for the second peak).}
 \label{fig2}
\end{figure}

\begin{figure}
\centering
 \resizebox{\hsize}{!}{\includegraphics{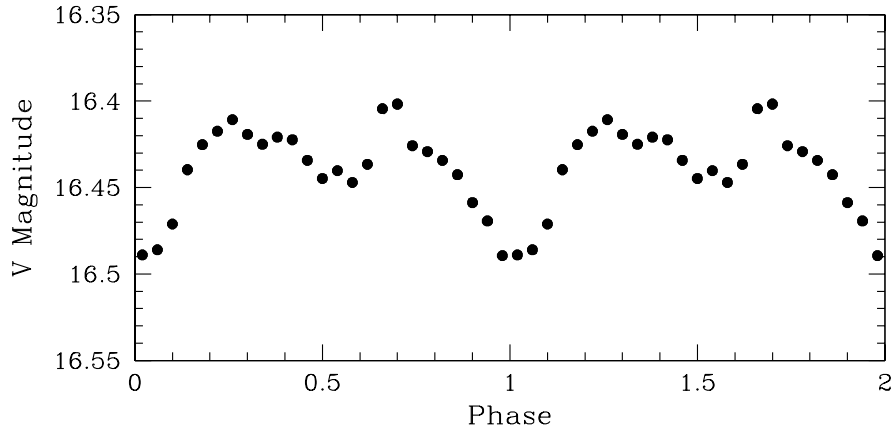}}
 \caption[]{The mean light curve of RXJ0944 (run S6324), folded on the 2162 s periodicity.}
 \label{fig3}
\end{figure}

It was obvious early in our work that RXJ0944 has a repetitive brightness modulation
with a period $\sim$ 2000 s. With further observations it could be seen that the 
feature is a double humped profile, with the two humps varying independently and rapidly 
in amplitude. In Figure~\ref{fig2} we show the light curve of run S6324 on a larger 
scale, with the cyclic modulation marked, and its highly variable pair of peaks. The FT for 
this run discloses a fundamental period at  $\sim$ 2220 s plus its first harmonic. There 
are only six cycles of this modulation in the light curve, so the uncertainty of the 
period is large (at least $\sim$ 40 s). The mean light curve, folded on the fundamental period
of 2162 s as derived below, is 
given in Figure~\ref{fig3} and shows the double humped nature of the profile, and that the humps sit on 
plateaux with only short-lived dips between them. (We removed the strong flare seen at HJD 2452356.418 
in Figure~\ref{fig2} as being not representative; it probably resulted from a sudden short-lived surge
of mass transference.) In the mean light curve, the two peaks occur at about phases
0.26 and 0.68, respectively.
    
\begin{figure}
\centering
 \resizebox{\hsize}{!}{\includegraphics{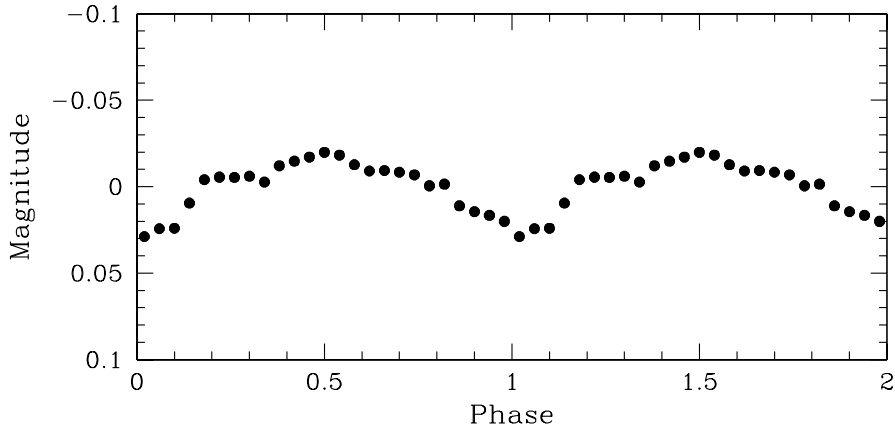}}
 \caption[]{The mean light curve of RXJ0944 (all observations), folded on the 2162 s periodicity.}
 \label{fig4}
\end{figure}

\begin{figure}
\centering
 \resizebox{\hsize}{!}{\includegraphics{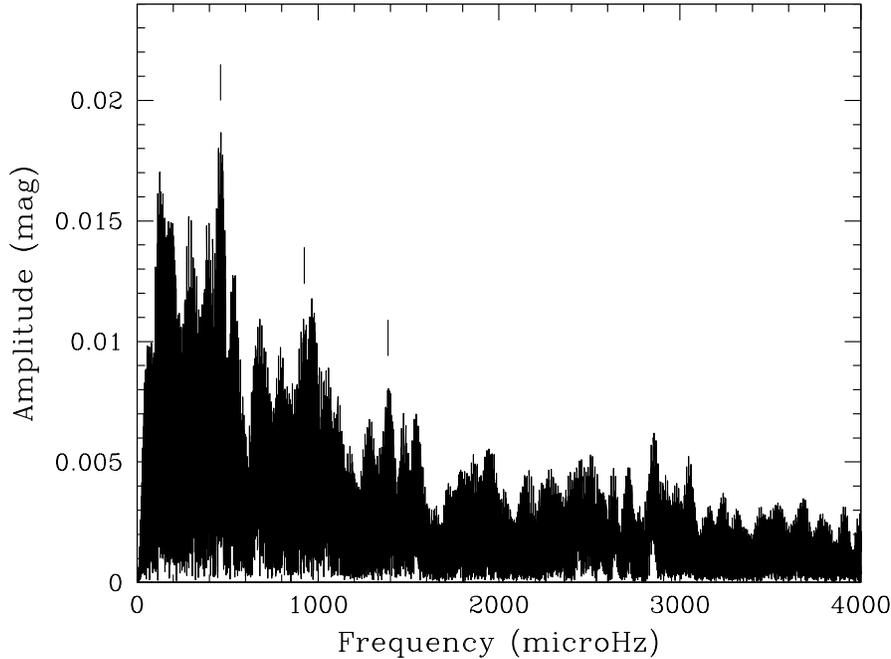}}
 \caption[]{The Fourier Transform of the complete data set of RXJ0944. The fundamental and the first two
harmonics of the 2162 s periodicity are marked by vertical bars.}
 \label{fig5}
\end{figure}

The peaks on the plateau appear as flares of variable width, so
that adding more observations tends to even out their contributions, with the result
that the mean light curve for the entire data set (using the period of 2162 s), 
shown in Figure~\ref{fig4}, has largely lost the evidence for the doubling of the profile. The FT for 
the full set of observations  is given in Figure~\ref{fig5}, and shows clearly the humps of power 
near the $\sim$ 2000 s fundamental and its first and second harmonics. There is a great deal of 
complicated fine structure in the FT, beyond what is produced by the window pattern; 
this is caused by the rapid amplitude modulation of the fundamental and its harmonics. It 
is not possible to select unambiguous frequencies from the forest of aliases. 
However, the highest peak in the neighbourhood of the fundamental modulation is at 2162 s 
and the highest peak at the first harmonic is 1079 s, which supports the choice of a 
fundamental period near 2160 s.
   
There are other humps of power in the total FT, but by subdividing our 
data (in particular, treating the March and April data sets separately) we find that 
the FT is non-stationary -- only the 2160 s modulation and its harmonics are persistent 
features. Given the high activity in the light curves (Figure~\ref{fig1}) it is not surprising that the 
FT is also very variable.
    
We find no evidence for rapid oscillations in brightness (Dwarf Nova 
Oscillations -- typically with periods in the range 5--50 s: see Warner 1995), but in run 
S6341 we find a Quasi-Periodic Oscillation (QPO; see Warner 1995) with a mean period of 351 
s and amplitude 0.013 mag. This is clearly seen in the light curve and maintains 
coherence for about 6 cycles between each major change of phase.

\section{Discussion}

     The presence of two distinct coherent periodicities in a CV is the 
recognised signature of an intermediate polar (IP) in which the non-orbital modulation is 
the spin period ($P_{sp}$) of the white dwarf primary, or its orbital side band (see, e.g., Warner 
1995). X-ray emission is another common feature of IPs, resulting from accretion from the 
inner edge of the accretion disc onto the magnetic pole(s) of the white dwarf. We 
therefore conclude that RXJ0944 is most probably an IP with highly variable two-pole accretion.
        
With $P_{orb}$ = 3.581 h and $P_{sp}$ = 36.0 min, RXJ0944 is quantitatively 
similar to canonical IPs such as FO Aqr and TV Col. However, the double-humped light 
curve and other properties make it most similar to YY Dra, as can be seen from the 
following brief review of the latter's properties.
    
YY Dra is a dwarf nova at V $\sim$ 16.0 quiescent magnitude with a mean 
outburst interval of 870 d and amplitude 5.5 mag, a $P_{orb}$  of 3.96 h and a 
$P_{sp}$ of 529.2 s. Both the spin
period and the orbital sideband (at 550 s) have been detected in the optical 
region (Patterson et al. 1992). YY Dra is the X-Ray source 3A1148+719 and the spin 
modulation is seen in the X-Ray emission (Patterson \& Szkody 1993). An M-type spectrum 
of the secondary is visible in the red, which is not normal for CVs with $P_{orb}$ $\sim$ 4h, 
suggesting a lower luminosity disc, probably the result if its central truncation by the 
magnetosphere of the primary.
    
HST observations of YY Dra (Haswell et al. 1997) show that the UV 
emission line profiles are modulated at half the spin period, and that there is 
simultaneous presence of broad red and blue wings in C\,IV emission, which is interpreted as evidence 
for two-pole accretion.  This is in accord with the double-humped pulse profiles in the 
optical and X-ray regions, and the occasional variations in height of the two peaks 
(though YY Dra is also notable for the near equality of its accretion pole luminosities most 
of the time).
During an outburst of YY Dra X-ray emission greatly increased and the 529 s 
oscillation was usually visible, but near maximum disappeared, which is interpreted as 
possible due to the equality of accretion onto two extended poles (Szkody et al. 2002).
    
There are other IPs with evidence of two-pole accretion: V405 Aur (Haberl et al. 1994)
is similar in having its principal optical modulation at the first harmonic 
rather than the 545 s fundamental (Allan et al. 1996); the 1WGA J1958.2+3232 ($P_{sp}$ = 1467 s) 
has a double peaked profile which shows reversal of circular polarization between 
the peaks, confirming that it is a two-pole accretor (Norton et al. 2002) and Still, 
Duck \& Marsh (1998) have found spectroscopic evidence that RX\, J0558+5353 is a two-pole 
accretor.
    
The basic similarity of the optical photometric properties of RXJ0944 and YY Dra is
evident, so we conclude that the same model, namely two-pole accretion, is 
the most probable description of RXJ0944, though with more variable and independent 
accretion rates onto the two poles. The phases of the two peaks determined from the mean 
light curve (Figure~\ref{fig3}) are not half a cycle apart, indicating that the magnetic poles
are not diametrically opposite on the surface of the primary.
For an IP, however, the strengths of the He\,II and 
C\,III/N\,III emission lines at 4686 {\AA} and 4650 {\AA}  (Mennickent et al.~2002) are relatively 
weak -- but the fact that they are seen at all is uncharacteristic of an ordinary dwarf 
nova.

\section{Conclusion}

RXJ0944 has the spectroscopic and photometric characteristics of an IP and 
would be worth studying in a pointed X-ray observation, in order to detect any 
modulation and thereby determine whether the 36.0 min periodicity in the optical is the 
white dwarf spin period or an orbital sideband. A time-resolved spectroscopic study with the 
HST should also be undertaken. RXJ0944 also has interest as a dwarf nova -- there are 
other intermediate polars that show full dwarf nova outbursts (XY Ari is an 
example) and others that have abbreviated outbursts (e.g. V1223 Sgr). High speed 
photometry during an outburst of RXJ0944 could help to reveal the interaction between disc and
magnetosphere as the rate of mass transfer increases and decreases.

\acknowledgements

We thank Drs S. Potter and P. Rodriguez-Gil for allowing us to use their 
light curve of
RXJ0944. PAW is supported by funds from the University of Cape Town and the
National Research Foundation; BW is supported by funds from the University.


\end{article}
\end{document}